\documentclass[letterpaper]{jpconf_marginfix}
\usepackage{graphicx}
\usepackage{bm}
\usepackage{eufrak}
\usepackage{slashed}
\usepackage{amsmath}
\usepackage{epsfig}
\newcommand{\exclude}[1]{}
\begin{document}
\title{Observational Consequences of Topological Currents in Neutron Stars}

\author{James Charbonneau}

\address{University of British Columbia, \em{james@phas.ubc.ca}}

\begin{abstract}
We argue that topological vector currents may be the source of many phenomena in neutron stars: kicks, jets, toroidal fields and magnetic helicity.  Topological vector currents exist because of the P-symmetry violation of the weak interaction.  Kicks and helicity are both objects that transform as pseudovectors and thus require P-symmetry violation to manifest themselves.  This symmetry argument is supported numerically; topological currents provide transfer enough momentum to describe even the largest of kicks and can generate large toroidal fields that create helicity. An observational signature of these currents is a faint left circular polarization in the X-rays in the wake of the neutron star that may require high precision polarimetry to see.  This paper is prepared from material presented at the 24th Lake Louise Winter Institute. 
\end{abstract}

\section{Introduction}
This paper distils some of the observational results originally presented in \cite{Charbonneau:2009ax}.  The goal of the aforementioned paper was to show that topological currents exist in neutron stars and to estimate their magnitude in different states of matter.  Here we reiterate some of the applications of topological currents to various neutron star phenomena buried in that paper: kicks, jets,  toroidal fields and magnetic helicity.   The use of topological currents as a possible resolution of the contradiction between precession and type-II superconductivity is detailed in \cite{Charbonneau:2007db}. Topological currents are well established (see  \cite{Charbonneau:2009ax} for references) and the vector current has the form, 
\begin{eqnarray}  
\label{j}
\langle j \rangle = (\mu_l - \mu_r)\frac{e \Phi}{2\pi^2},
\end{eqnarray}  
where $\mu_r$ and $\mu_l$ are the chemical potential of the right and left-handed electrons, and $\Phi$ is the magnetic flux.  Because of their topological nature these currents are non-dissipative.

Equation  (\ref{j}) has a very simple physical meaning: to compute the current one simply counts the difference between left-handed and right-handed electrons in the background of a magnetic field.  In a magnetic field the spin of an electron will tend to antialign with the field.  Then, if there is an excess of left-handed electrons they must move in the direction of the magnetic field and if there is an excess of right-handed electrons they must move against the field.  The topological vector current acts as a pump to remove the average helicity and it stops pumping once the average helicity is zero again.   

There are three requirements for topological vector currents to be present: an imbalance in left and right-handed particles $\mu_l\neq \mu_r$, degenerate matter $\mu\gg T$, and the presence of the background magnetic field $B\neq 0$.  All of these  are present in neutron and quark stars.  The weak interaction, by which the star attains equilibrium, violates parity; particles created in this environment are primarily left-handed.  The interior of the star is very dense, $\mu_e = 100$ MeV, and cold, $T=0.1$ MeV, such that the degeneracy condition $\mu\gg T$ is met, and neutron stars are known to have  a huge magnetic field, $B\sim10^{12}$ G.  

There is a subtlety to consider. In an infinite system any asymmetry in left and right-handed electrons created by the weak interaction would be washed out; the creation and annihilation rates of the left-handed particles are the same.\footnote{We neglect all QED re-scattering effects.  Though they are much stronger than weak interactions they are P-even and cannot wash out the asymmetry.  Because of the large magnetic field the electron only travels in the direction of the magnetic field while the motion in transverse directions confined to Landau levels. }  Though many more left-handed electrons are created, they are also destroyed much faster than the right-handed and no asymmetry builds.  The key is that the neutron star is a finite system and electrons are removed from the star by the topological current before they can decay.  This topological current corresponds to the lowest energy state in the thermodynamic equilibrium when $ \mu_l \neq \mu_r$ is held fixed. In reality there is a tendency for $ \mu_l $ and $\mu_r$ to equilibrate though weak interactions; however, due to the finite size of the system a complete equilibrium can not be achieved. 

The current allows electrons to be carried out of the system, but the electron chemical potential does not drop to zero as it does for neutrinos. The induced current only remains  non-dissipating when the system is degenerate, $\mu\gg T$.  In the star's crust this condition becomes invalid, the current will become dissipative, and the trapped electrons will return into the system.  If the electrons manage to escape the star the electron chemical potential will slowly decrease until the current stops.  Charge neutrality will cause matter to accrete isotropically possibly maintaining some of the chemical potential. 

Equation  (\ref{j}) presents a simplified picture and is not sufficient to estimate the current in a neutron star.  It only accounts for the lowest landau level at zero temperature.  Obtaining a good estimate is quite cumbersome and the reader is encouraged to read  \cite{Charbonneau:2009ax} for details.  The phase and composition of matter is hotly debated and to simplify the discussion we will provide a typical value for the current,
\begin{eqnarray}
\label{j_estimate}
\langle j \rangle \sim 10^{-10}  \, \left(\frac{T}{10^{9} \textrm{ K}}\right)^5  \textrm{ MeV \quad   or  \quad   } e \langle j\rangle \sim   10^{-8}  \, \left(\frac{T}{10^{9} \textrm{ K}}\right)^5  \textrm{ A.}
\end{eqnarray}
This value and the temperature dependence reasonably describe the current in quark stars, neutron stars with kaon condensates and neutron stars with hyperons or large proton fractions.

\section{Applications of topological currents in neutron stars}
\subsection{Neutron star kicks}
We first consider a well established phenomenon -- neutron star kicks.  It is accepted that neutron stars have much higher proper velocities than their progenitors.  Large velocities are unambiguously confirmed with the model independent measurement of pulsar B1508+55 moving at $1083^{+103}_{-90}$ km/s \cite{Chatterjee:2005mj}. There also appears to be a strong correlation between the direction of the kick and the spin axis of the star \cite{Wang:2006zia}.  Currently no mechanism exists that can reliably kick the star hard enough.  Asymmetric explosions can only reach $200$ km/s \cite{Fryer:2003tc}, and asymmetric neutrino emission is plagued by the problem that at temperatures high enough to produce the kick the neutrino is trapped inside the star \cite{Sagert:2007ug}. 

If the electrons carried by the current can transfer their momentum into space -- either by being ejected or by radiating photons -- the current could push the star like a rocket.   In typical neutron stars this is unlikely because the crust (the region where $\mu\sim T$) is thought to be very thick.  But if the crust is very narrow, as it may be in bare quark stars \cite{Alcock:1986hz} (see recent developments \cite{Alford:2006bx}), the electrons may leave the system or emit photons that will carry their momentum to space.   With this in mind we conjecture  that stars with very large kicks, $v\gg 200$ km/s, are quark stars and that slow moving stars, $v \leq 200$ km/s, are kicked by some other means  \cite{Fryer:2003tc} and are typical neutron stars.  Confirmation of this would provide an elegant way to discriminate between neutron stars and quark stars.

Topological kicks work by slowly and steadily pushing the star over time, much like a rocket.  Even if the magnetic field and spin axis are not perfectly aligned, the kick will align with the spin axis because the spin period is much larger than the kick duration.  This is contrary to other mechanisms where the kicks occur shortly after the star's birth.  In reality the neutron star cools as it ages, but \cite{Charbonneau:2009ax} presents a conservative estimate of the time required to push a star to $1000$ km/s by choosing a temperature when the star has cooled,
\begin{eqnarray}
t = 10^{12} \left(\frac{v}{1000 \textrm{ km/s}}\right) \left(\frac{B^{12}{\textrm{ G}}}{B}\right)\left(\frac{10^{9} \textrm{ K}}{T}\right)^5 \textrm{s}
\approx 40,000 \textrm{ years,}
\end{eqnarray}
which is on the order of neutron star ages.  The current is much larger at the star's birth, yielding larger kicks, and can easily explain velocities as seen, for example, in the Vela pulsar.  

The correlation between the kick velocity and the magnetic field $\langle  \vec{v}\cdot \vec{B} \rangle$ is  P-odd and indicates that it must be generated by a parity violating process.  An important aspect of our  kick engine is that it  continues to work at low temperatures $T \ll$ MeV, unlike neutrino kicks.  This is because it is the chemical potential that drives the kick, not the temperature.

\subsection{Toroidal magnetic fields}
There is strong theoretical evidence for the existence of toroidal fields in neutron stars  based on the stability of the poloidal magnetic field.  Toroidal and poloidal fields of similar magnitudes must be necessary to stave off hydrodynamic instabilities -- the toroidal field suppresses poloidal instabilities and vice versa, for example see \cite{Markey:1973}.  A toroidal field is a natural consequence of having a current running along the poloidal field.

Estimating the toroidal magnetic field is a very complicated problem that requires
a self consistent solution of the equation of the magnetic hydrodynamics.  Our induced, topological currents represent  only a small part of the system. We are not attempting to solve this problem. Instead, we shall argue that the currents we estimated are more than sufficient to induce a toroidal magnetic field correlated on the scale of neutron stars, $R=10$ km.

The calculation of the magnetic field is complicated by the fact we are inducing a magnetic field inside a superconductor.  An estimate outlined in \cite{Charbonneau:2009ax} using Ampere's law leads to the following expression for $H_{\textrm{tor}}$ in terms of   magnitude of poloidal magnetic field $H$, 
\begin{eqnarray}
\label{Ampere-1}
 \frac{H_{tor}}{H}  \sim  4  \left(\frac{\langle j\rangle}{10^{-10}\text{ MeV}}\right)
 \left(\frac{L}{ \text{km}}\right).
\end{eqnarray}
This shows that a typical current (\ref{j_estimate}) can induce a toroidal field the same magnitude as the poloidal field on scales the order $L\sim 1$ km, within the typical size of a neutron star.  Our estimate becomes unreliable when $H_{\textrm{tor}}\geq H$ and we can no longer ignore the current induced by the toroidal field and the problem  requires a self consistent analysis.

\subsection{Magnetic helicity}
Our estimate for the induced toroidal field $H_{\textrm{tor}}$ unambiguously implies that the magnetic helicity will be also induced.  Magnetic helicity $\mathcal{H}\equiv\int \!d^3x \,\vec{A}\cdot\vec{B}$ is a topological quantity that measures the linking of field lines.   In a magneto-fluid with zero resistivity, a very good approximation for neutron stars, it becomes a topological invariant.  This invariance also provides a stability mechanism for the poloidal field.  Magnetic helicity  is the dot product of a vector and a pseudovector, making it a pseudoscalar.  Under the parity transformation $\vec{x}\rightarrow -\vec{x}$ the magnetic helicity is P-odd:  $\mathcal{H}\rightarrow -\mathcal{H}$.  This implies that  the magnetic helicity can be only induced if there are P-symmetry violating processes producing   a  large coherent effect on macroscopic scales.  Many attempts to generate helicity rely on instabilities in the magnetic field caused by the star's rotation.  Such correlations $\vec{B}\cdot\vec{\Omega}$ are P-even, and though they may generate toroidal fields they cannot be responsible for helicity.

\subsection{Pulsar jets and X-ray polarimetry}
It has been argued that spin axis and the direction of the proper motion of the Crab and Vela pulsars are aligned and that the pulsar jets \cite{1538-4357-554-2-L189} are correlated neutron star kicks \cite{Wang:2005jg}.  The current, and thus the proper motion, is aligned with the magnetic field, which itself is correlated with the axis of rotation.  It is  tempting to identify the observed inner jets \cite{1538-4357-554-2-L189} with the electrons/photons emitted as a result of the induced current.    

It is possible that evidence of the topological current may be directly detected in these jets. An observational consequence of the current is that a component of the X-ray emission in the trail of the neutron star will be left circularly polarized.  Because of the parity violation in the star, only left-handed electrons will contribute to the kick.  When these left-handed electrons interact they will create mostly left-handed photons, which coherently will be seen as left circularly polarized X-rays.  As there are many other sources of X-rays the contribution is likely very small, but it further motivates the need for higher precision X-ray polarimetry  \cite{Weisskopf:2006xb}.   
 
\ack
This work was supported by the National Science and Engineering Council of Canada.

\section*{References}
\providecommand{\newblock}{}

\end{document}